\documentclass[a4paper]{spie}  

\usepackage{amsmath,amsfonts,amssymb}
\usepackage[dvipsnames]{xcolor}
\usepackage{graphicx}
\usepackage[inkscapelatex=false]{svg}
\usepackage{subcaption}
\usepackage{xspace}

\usepackage{tikz}
\usetikzlibrary{arrows.meta,positioning,shapes.geometric,calc,fit}

\definecolor{arrowpurple}{RGB}{126,87,194}
\definecolor{envblue}{RGB}{45,120,180}
\definecolor{agentred}{RGB}{200,65,65}
\definecolor{obsgreen}{RGB}{85, 145, 120}
\usepackage{}
\tikzset{
  block/.style = {draw, rounded corners, align=center, minimum width=2.8cm, minimum height=1.0cm},
  smallblock/.style = {draw, rounded corners, align=center, minimum width=2.4cm, minimum height=0.9cm},
  data/.style = {draw, align=center, minimum width=3.2cm, minimum height=1.0cm, fill=gray!10},
  line/.style = {->, >=Latex, very thick, draw=arrowpurple},
  dashedline/.style = {->, >=Latex, very thick, dashed, draw=purple},
  agent/.style = {block, draw=agentred, very thick},
  obs/.style = {block, draw=obsgreen, very thick},
  environment/.style = {draw=envblue, dashed, rounded corners, thick, inner sep=8pt},
  statevar/.style={text=red, font=\large\bfseries}
}

\usepackage{xparse} 
\usepackage[normalem]{ulem} 
\usepackage{soulutf8} 

\usepackage[colorlinks=true, allcolors=blue]{hyperref}

\NewDocumentCommand{\manote}{O{}m}{\textbf{[Note #1: #2]}}

\newcommand{\ie}{\emph{i.e.},\xspace}

\usepackage[a4paper,left=1.90cm,right=1.9cm,top=2.6cm,bottom=5cm]{geometry}

\title{Reinforcement learning for post-coronagraphic wavefront control}

\author[a,b,c]{Manuela Castañeda-Medina}
\author[a,b,c]{Yann Gutierrez}
\author[c]{Johan Mazoyer}
\author[a]{Baptiste Abeloos}
\author[b]{Laurent M.\ Mugnier}
\author[a]{Olivier Herscovici-Schiller}
\affil[a]{DTIS, ONERA, Université Paris-Saclay, 91120 Palaiseau, France} 
\affil[b]{DOTA, ONERA, Université Paris Saclay, 92322 Châtillon, France} 
\affil[c]{LIRA, Observatoire de Paris, Université PSL, Sorbonne Université, Université Paris Cité,
CNRS, 5 place Jules Janssen, 92195 Meudon, France} 

\authorinfo{Send correspondence to laura\_manuela.castaneda\_medina@onera.fr}

\begin{document} 
\maketitle

\begin{abstract}
Direct imaging of exoplanets is limited by the extreme contrast between the star and the planets, which is mitigated using a coronagraph. However, optical aberrations cause starlight leakage through the coronagraph, producing speckles that obscure the planetary signal. Achieving the required contrast levels demands wavefront control with subnanometric precision. Deep reinforcement learning offers a promising alternative to traditional focal-plane wavefront control techniques by enabling adaptive correction strategies learned directly from interaction with the system. In this work, we present a fully data-driven method for post-coronagraphic aberration correction in a simulated high-contrast imaging testbed. The agent controls a deformable mirror using observations consisting of focal-plane measurements (images) and physics-informed wavefront sensing information derived from these images. We evaluate different observation representations and control strategies, and the method is validated on simplified simulations of a high-contrast imaging testbed, where it successfully creates dark holes, \ie  regions of the focal plane in which residual starlight is strongly suppressed, while approaching the performance of conventional wavefront control methods.
\end{abstract}


\section{INTRODUCTION}
\label{sec:intro}

To date, more than 6,000 exoplanets have been detected using a variety of observational techniques. Direct imaging aims to capture photons emitted or reflected by the planet itself \cite{Currie2023DirectImagingSpectroscopy}, enabling spectroscopic analysis of exoplanet atmospheres and the detection of molecular signatures related to their composition and physical conditions. However, direct imaging remains extremely challenging because exoplanets can be up to ten orders of magnitude fainter than their host stars and are observed at small angular separations. Coronagraphs suppress the stellar core and diffracted light, but optical aberrations introduced along optical paths not shared by the telescope's sensing and science channels, known as non-common-path aberrations (NCPAs), cause residual starlight to leak through the coronagraph. Slowly evolving aberrations produce quasi-static speckles in the focal plane that can mimic or obscure planetary signals. Active wavefront sensing and control is therefore required to estimate and correct these aberrations using deformable mirrors, creating a dark hole, \emph{i.e.}, a region in the focal plane where residual starlight is strongly suppressed \cite{GalicherMazoyer2023}.

Although these challenges concern both ground- and space-based observatories, this work focuses on space-based applications. The Coronagraph Instrument aboard the Nancy Grace Roman Space Telescope will provide an in-space demonstration of active coronagraphy and wavefront control \cite{Cady2025Roman}, paving the way for the Habitable Worlds Observatory (HWO), which aims to directly image and characterize Earth-like exoplanets around nearby Sun-like stars \cite{Feinberg2026HWO}. Classical approaches, such as Pair-Wise Probing combined with Electric Field Conjugation (PWP+EFC), rely on a linearized optical model to estimate and cancel speckles and constitute a state-of-the-art baseline for high-contrast coronagraphy \cite{Giveon2007EFC}. These methods perform well under accurate calibration but can degrade under model mismatch, temporal drifts, or measurement noise.

In this work, dark-hole generation is framed as a sequential decision problem solved with reinforcement learning (RL) \cite{SuttonBarto2018RL}. The primary objective is to investigate model-free wavefront control, in which the agent learns the correction strategy rather than computing each DM command by explicitly inverting a linearized optical model. In the present implementation, however, the policy observations still include model-dependent information obtained through PWP and EFC. This approach aims to reduce the number of correction iterations and, when possible, improve the final contrast, while exploring robust strategies for future space telescopes dedicated to exoplanet characterization.

\section{METHOD}

\subsection{Reinforcement Learning}

Reinforcement learning (RL) is a framework where an agent learns to make decisions by interacting with an environment in order to maximize a reward signal \cite{SuttonBarto2018RL}. RL problems are commonly modeled as a \emph{Markov decision process}, defined by
\begin{equation}
(\mathcal{S}, \mathcal{A}, P, R, \gamma),
\end{equation}
where $\mathcal{S}$ is the state space, $\mathcal{A}$ the action space, $P(s'|s,a)$ the transition probability, $R(s,a)$ the reward function, and $\gamma \in [0,1]$ the discount factor. The agent’s behavior is described by a \emph{policy} $\pi(a|s)$, which is optimized to maximize the expected return.

Reinforcement learning has previously been investigated for adaptive optics control, particularly to address temporal delays, calibration errors, and nonlinearities in ground-based systems. Nousiainen et al.\ developed a model-based RL controller for atmospheric turbulence correction and demonstrated it in both numerical simulations and laboratory experiments \cite{Nousiainen2021AO,Nousiainen2022AO,Nousiainen2024Laboratory}. In parallel, Gutiérrez et al.\ introduced a model-free, image-based approach in which an RL agent learns to estimate and correct non-coronagraphic wavefront aberrations directly from focal-plane phase-diversity images \cite{Gutierrez2024ImageBased}.

This approach was subsequently extended to post-coronagraphic wavefront control by Gutiérrez et al.\ \cite{Gutierrez2024DeepRL}, demonstrating the potential of RL agents to control a deformable mirror and create a dark hole from focal-plane measurements. Building on this work, we investigate model-free RL for dark-hole creation in a simplified coronagraphic simulation. We focus on the design of the observations and reward function, compare the resulting controller with PWP+EFC, and evaluate its initial robustness to photon noise.

In our approach, wavefront correction is framed as an RL problem in which the environment represents the coronagraphic optical system, the actions correspond to commands applied to the DM actuators, and the observations combine focal-plane images with physics-informed wavefront sensing and control information derived from these images. The reward is computed from the residual intensity, or contrast, within the dark hole. At each interaction step, the agent proposes a DM command, which is propagated through the optical system to produce a new focal-plane image. The resulting observation and reward define a transition that is used to train the policy.

\subsection{RL-based wavefront control}

Wavefront correction is formulated as an RL problem in which the environment represents the coronagraphic optical system. An episode begins with a newly generated static entrance aberration and consists of five correction steps. At each step, the agent receives an observation, proposes an incremental command for the deformable mirror, and observes the resulting coronagraphic focal-plane measurements.

The observation contains four complementary components: the current science image, the complex electric field estimated through pair-wise probing, the change in the estimated electric field with respect to the preceding step, and the correction proposed by EFC. The science image provides a direct measurement of the residual intensity, whereas the electric-field estimate contains phase and amplitude information that cannot be recovered from a single intensity image. The electric-field variation provides information about the response of the optical system to the preceding action, while the EFC command supplies a physics-informed correction based on the synthetic optical model.

\begin{figure}[ht]
  \centering
  \begin{tikzpicture}[scale=0.9, transform shape, node distance=1cm]

    \node[block] (telescope) {Telescope};
    \node[block, below=1.6cm of telescope] (dm) {Deformable\\mirror};
    \node[agent, left=4.5cm of dm] (agent) {Agent};
    \node[block, right=2cm of dm] (corona) {Coronagraph};
    \node[block, below=3cm of dm] (detector) {Detector};
    \node[obs, below=3cm of agent] (imgs) {Data pre-processing\\(Observation)};

    \draw[line] (telescope) -- node[right]{Aberrations} (dm);
    \draw[line] (dm) -- (corona);
    \draw[line] (agent) -- node[above, statevar] {$A_t$} (dm);
    \draw[line] (agent) -- node[below]{\small (DM actuator commands)} (dm);
    \draw[line] (imgs) -- node[right, statevar] {$O_t$} (agent);
    \draw[line] (imgs) -- ++(-4,0) coordinate (c1) -- node[left, statevar] {$R_t$} (c1 |- agent.west) -- (agent.west);
    \draw[line] (imgs) -- ++(-4,0) coordinate (c1) -- node[right]{} (c1 |- agent.west) -- (agent.west);
    \draw[line] (detector) -- (imgs);
    \draw[line] (corona) |- (detector.east);

    \node[environment,
      fit=(telescope) (dm) (corona) (detector),
      label={[text=blue, align=center]south:Environment}] (envbox) {};

  \end{tikzpicture}
  \caption{RL pipeline in wavefront control}
  \label{fig:rlwf_diagram}
\end{figure}
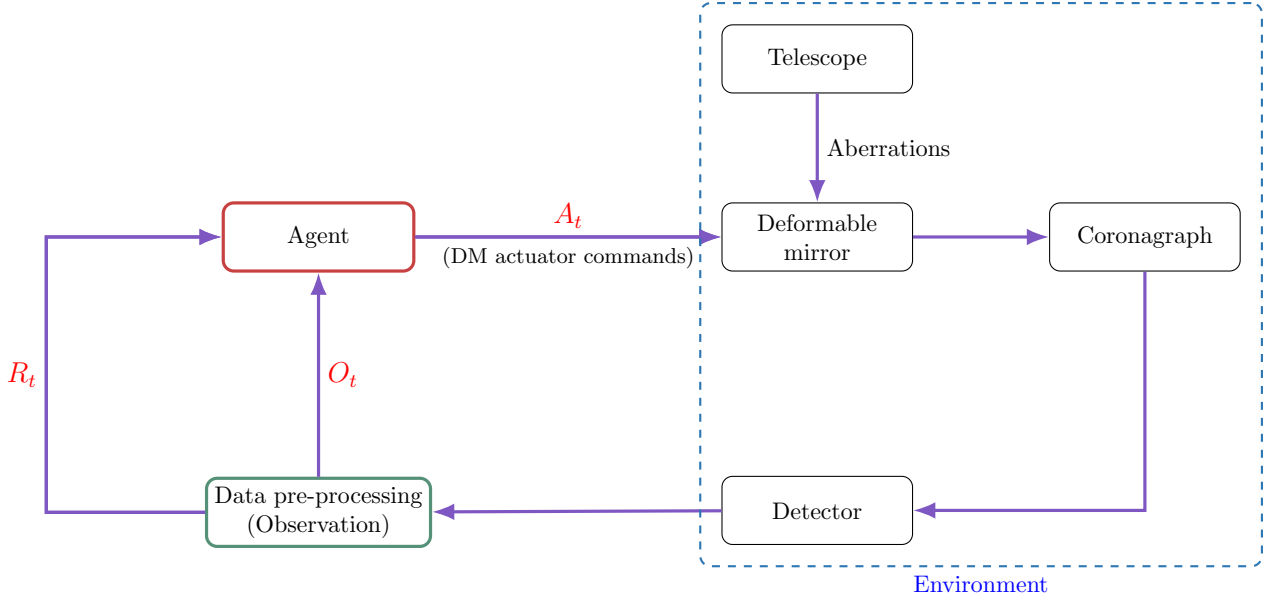

The action is a continuous vector containing one incremental command for each controlled DM actuator. The policy output is bounded to $[-1,1]$ and scaled by the environment before being added to the current DM shape. In the configuration considered here, the action space therefore contains 62 controlled degrees of freedom. Finally, the reward encourages both the attainment of deep contrast and the improvement obtained during the current correction step. Denoting by $C_t$ the mean contrast inside the dark hole after step $t$, we studied several contrast-based metrics and found that the following reward gave the best results:
\begin{equation}
    R_t =
    \left[-\log_{10}\left(C_t\right)\right]
    \log_{10}\left(
        \frac{C_{t-1}}{C_t}
    \right).
\end{equation}
The first factor favors low absolute contrast, while the second rewards an improvement relative to the preceding step. Consequently, an action that reduces the dark-hole contrast receives a positive reward, whereas an action that degrades it receives a negative reward.

\subsection{PWP + EFC}

Pair-wise probing combined with Electric Field Conjugation (PWP+EFC) is a classical wavefront sensing and control approach widely used for high-contrast coronagraphic imaging. Because non-common-path aberrations must be sensed at the science detector itself, PWP estimates the complex electric field directly in the coronagraphic focal plane rather than relying on a separate wavefront sensor. Its objective is to estimate this field and compute the deformable mirror commands required to suppress residual starlight within the dark hole.

The first step, pair-wise probing, estimates the complex electric field associated with the speckle pattern. A synthetic linear model of the optical system is used to predict the focal-plane electric field generated by a known DM probe pattern. Each probe is applied with both positive and negative signs, producing the intensity images \(I_k^{\pm}\), which can be modeled, within a first-order Taylor expansion, by
\begin{equation}
    I_k^{\pm} = \left|E \pm iC[A\phi_k]\right|^2,
\end{equation}
where \(E\) is the unknown focal-plane electric field, \(A\) is the unaberrated pupil-plane electric field, \(\phi_k\) is the phase introduced by the \(k\)-th DM probe, and \(C\) denotes propagation from the pupil plane to the coronagraphic focal plane. Hence, \(iC[A\phi_k]\) is the modeled focal-plane field introduced by the probe. Subtracting the two images cancels the terms that are unchanged between the exposures, including the unprobed speckle intensity and the probe intensity, yielding
\begin{equation}
    I_k^{+}-I_k^{-}
    =
    4\left(
    \mathrm{Re}\left(E\right)\mathrm{Re}\left(iC[A\phi_k]\right)
    +
    \mathrm{Im}\left(E\right)\mathrm{Im}\left(iC[A\phi_k]\right)
    \right).
\end{equation}
Thus, the intensity difference isolates the interference between the unknown speckle field and the known probe field. Because a detector measures intensity rather than the complex field directly, at least two probe pairs producing linearly independent focal-plane modulations are required to recover the real and imaginary components of \(E\) at each pixel. The resulting linear system is inverted using the synthetic probe model to obtain the electric-field estimate. A detailed description of the PWP formalism and its implementation is provided by Potier et al.\ \cite{Potier2020}.

Once the electric field has been estimated, Electric Field Conjugation computes the DM command that minimizes the residual intensity within the dark hole. A synthetic linear model of the optical system is used to construct the interaction matrix \(G\), which describes the linearly approximated change in the focal-plane electric field produced by each DM actuator. Under this linear approximation, the corrected electric field is written as
\begin{equation}
    E_{\mathrm{c}} = \widehat{E} + Gu,
    \label{eq:efc_correction}
\end{equation}
where \(\widehat{E}\) is the estimated electric field, \(G\) is the interaction matrix, and \(u\) is the vector of DM commands. Equation~\eqref{eq:efc_correction} can be minimized using different inverse-problem strategies. In EFC, singular value decomposition (SVD) is used to compute a regularized pseudoinverse of \(G\), yielding the control matrix \(G^{\dagger}\). Regularization limits the contribution of poorly constrained modes and prevents noise amplification. A detailed derivation of the EFC formalism, as well as a discussion of other inversion approaches, is provided by Groff et al.~\cite{Groff2016FocalPlane}.

At each iteration, PWP estimates the current electric field, EFC computes the corresponding correction, and the DM is updated. PWP+EFC has demonstrated strong performance under accurate calibration and stable conditions. It is part of the high-order wavefront sensing and control architecture developed and tested for the Roman Coronagraph Instrument \cite{Cady2025Roman} and has been demonstrated on sky with VLT/SPHERE \cite{Potier2022SPHERE}, making it a reference method for model-based focal-plane wavefront control.

However, PWP+EFC performance depends on the validity of two linear approximations and on the accuracy of the synthetic optical model. First, PWP relies on a first-order approximation of the field introduced by the probes; at large probe amplitudes, neglected higher-order terms can bias the electric-field estimate. Second, EFC relies on a linearized relationship between DM commands and changes in the focal-plane electric field; at large DM strokes, this approximation can become inaccurate and limit loop convergence \cite{Laginja2025ExtendedLinearity}. Even within these linear regimes, discrepancies between the model and the real instrument can produce inaccurate electric-field estimates and suboptimal EFC commands \cite{Zhou2020RomanModelValidation}.

\section{EXPERIMENTAL SETUP}

\subsection{Optical simulation}

All experiments in this work were performed using the Asterix simulator, a Python-based library for high-contrast imaging simulations \cite{MazoyerAsterixSimulator}. The simulated monochromatic wavelength was $\lambda_0=500\,\mathrm{nm}$. The entrance pupil was circular and unobstructed, sampled on an 80-pixel-diameter pupil plane, with the actuator-grid configuration setting the effective pupil diameter to $2.4\,\mathrm{mm}$. The science detector contained $48\times48$ pixels sampled at 3 pixels per $\lambda/D$, corresponding to a $16\,\lambda/D$ field of view.

The coronagraph was enabled with an FQPM focal-plane mask, and performance was evaluated in a full square dark hole spanning $0$--$4\,\lambda/D$. Each episode used a newly generated static phase-only entrance aberration, represented by a phase screen with RMS phase $\sigma_\phi=0.1\,\mathrm{rad}$. The controller acted on the valid pupil actuators of an $8\times8$ DM3 actuator grid; after pupil thresholding, this yielded 62 controlled degrees of freedom.

To assess the noise-robustness of our method, we performed a set of experiments by introducing photon noise (described in Section 4.2). In these simplified experiments, the goal was to adjust the number of photon so that the SNR of the speckles remains at a constant over the whole correction. The requested SNR was defined as the mean Poisson SNR over the dark hole:
\begin{equation}
    {\rm SNR}_{\rm DH}
    =
    \left\langle
        \sqrt{I_i \beta N_{\rm tel}}
    \right\rangle_{i\in{\rm DH}},
\end{equation}
where $I_i$ is the noiseless contrast-normalized intensity in dark-hole pixel $i$, $N_{\rm tel}$ is the number of photons entering the telescope during an exposure and $\beta$ is the conversion factor between contrast-normalized intensity and the fraction of entering telescope photons expected in a detector pixel. Thus, the expected photon count in pixel $i$ is
\begin{equation}
    \mu_i = I_i \beta N_{\rm tel}.
\end{equation}
At reset, $N_{\rm tel}$ was calibrated from the initial dark-hole image to obtain the target value of ${\rm SNR}_{\rm DH}$. After each correction step, it was updated according to
\begin{equation}
    N_{\rm tel}(t)
    =
    N_{\rm tel}(0)
    \frac{C_{\rm DH}(0)}{C_{\rm DH}(t)},
    \label{eq:ntel_exposure}
\end{equation}
where $C_{\rm DH}(t)$ is the mean dark-hole contrast at step $t$. 

Equation~\eqref{eq:ntel_exposure} means that, in order to simulate an experiment where the mean dark-hole SNR remains approximately constant, the effective photon count is linearly increased as the contrast decreases. In a real experiment with a fixed-magnitude source, this would be achieved by linearly increasing the exposure time.

Photon noise was then generated by drawing
\begin{equation}
    n_i \sim {\rm Poisson}\!\left(I_i\beta N_{\rm tel}\right)
\end{equation}
and converting the resulting photon counts back to contrast units by dividing by $\beta N_{\rm tel}$.

\subsection{Training and Evaluation Setup}

At each environment reset, an entrance-pupil aberration was independently generated as a zero-mean Gaussian random phase screen with spatial power spectral density
\begin{equation}
    \mathrm{PSD}(\rho)\propto
    \left[1+\left(\frac{\rho}{\rho_c}\right)^2\right]^{-1},
    \qquad \rho_c=4.3,
\end{equation}
where
\begin{equation}
    \rho=\sqrt{k_x^2+k_y^2},
\end{equation}
\(k_x\) and \(k_y\) are the discrete Fourier-frequency indices of the actuator grid. Thus, \(\rho\) and the characteristic frequency \(\rho_c\) are expressed in FFT bins. If the \(8\times8\) actuator grid is assumed to span one pupil diameter, \(\rho_c=4.3\) corresponds approximately to \(4.3\) cycles per pupil, or to a characteristic spatial scale of \(D/4.3\simeq0.23D\). The phase screen was synthesized in the \(8\times8\) actuator influence-function basis, with piston removed, and rescaled to an RMS phase of \(0.1\,\mathrm{rad}\) over the entrance pupil. Each realization remained static throughout the corresponding episode. Training episodes contained at most five control steps, with early termination enabled.

Two pair-wise probe patterns were used, each applied with positive and negative signs. The amplitude of each signed probe command was \(20\,\mathrm{nm}\) in optical path difference, corresponding to a \(40\,\mathrm{nm}\) difference between the positive and negative commands. The policy generated incremental commands for the 62 active DM actuators located within the entrance pupil. Training was allowed to run for a maximum of \(5\times10^7\) environment interactions but could be stopped earlier once convergence was observed. The results reported here were obtained using the best saved policy checkpoint, selected during training, at \(6.6\times10^6\) environment interactions.

The selected policy was evaluated without further parameter updates. Actions were generated deterministically, with no explicit action noise, and generalized state-dependent exploration was disabled. The evaluation comprised 100 independently sampled episodes, each initialized with an aberration drawn from the same distribution as during training and held static throughout the episode. Episode length was set at 5 steps, for both training and evaluation. Performance was quantified by the arithmetic mean of the normalized detector intensity over the pixels within the dark hole. The initial uncorrected contrast and the contrast after each correction step were recorded. At every step, the distribution across the 100 episodes was summarized by its median and 25th--75th percentile interval.

Once the RL configuration had been selected, its performance was compared with the PWP+EFC reference. For each of the 100 comparison episodes, the RL and PWP+EFC environments were reset using the same seed, producing matched aberration realizations for the two controllers. Both methods used the same optical configuration and were evaluated over five correction steps. Here, one correction step denotes one policy action for the RL controller and one EFC correction for PWP+EFC. In both cases, the electric-field estimate used at each step was obtained from two probe pairs, requiring four probed detector images, in addition to the unprobed science image.

\section{RESULTS}

\subsection{Dark-Hole Creation with RL}

\begin{figure}[h!]
    \centering
    \includegraphics[width=\linewidth]{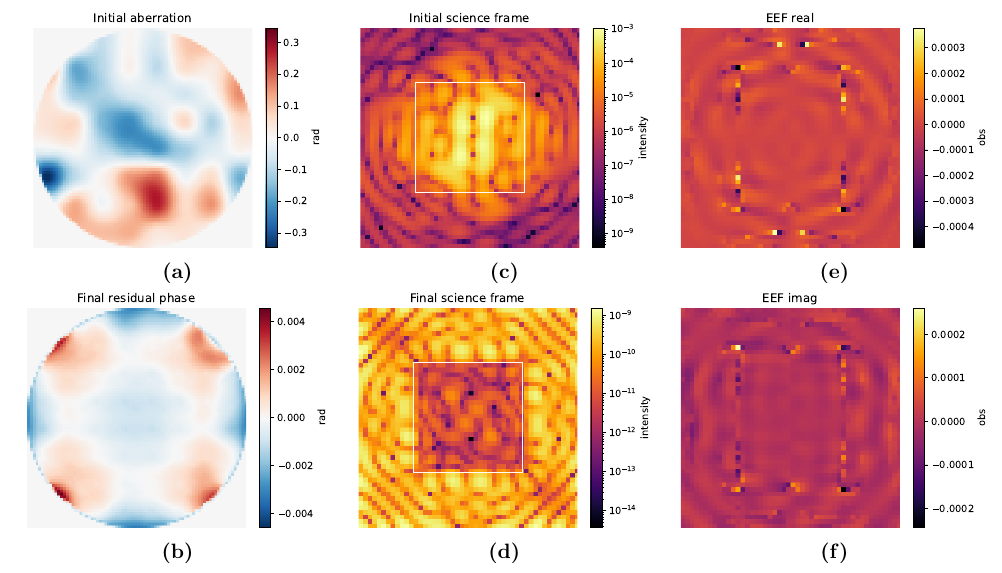}
    \caption{Initial and final optical states during dark-hole generation.
    (a) Initial entrance aberration at the start of the episode.
    (b) Final residual phase after correction.
    (c) Initial science frame before correction.
    (d) Final science frame after the RL control sequence.
    (e) Real part of the residual estimated electric field (\^E) after five control steps.
    (f) Imaginary part of the residual estimated electric field (\^E) after five control steps.}
    \label{fig:dark_hole_generation_states}
\end{figure}

Figure~\ref{fig:dark_hole_generation_states} shows the initial and final optical states during one representative episode of five control steps. Figure~\ref{fig:dark_hole_generation_states}(a) shows the initial entrance aberration at the start of the episode, while Fig.~\ref{fig:dark_hole_generation_states}(b) shows the final residual phase after correction. The residual phase exhibits the spatial structure imposed by the deformable-mirror correction applied to suppress the stellar residuals in the controlled region.

Figures~\ref{fig:dark_hole_generation_states}(c) and~\ref{fig:dark_hole_generation_states}(d) show the corresponding science frames before and after correction. The speckle field is initially distributed inside the dark-hole region, leading to a high residual intensity. After the RL control sequence, the final science frame shows the resulting dark hole, where the intensity has been strongly reduced with respect to the initial image.

Figures~\ref{fig:dark_hole_generation_states}(e) and~\ref{fig:dark_hole_generation_states}(f) show the real and imaginary parts of the residual estimated electric field after five control steps. The visible pixelation reflects imperfections in the PWP estimation in regions where the probes do not sufficiently modulate the electric field, making the corresponding estimation system poorly conditioned. The residual estimated electric field presents the dark-hole geometry, with a significantly reduced magnitude inside the controlled region. These qualitative results show that the learned policy is able to generate a dark hole within a short correction sequence.

\begin{figure}[h!]
    \centering
    \includegraphics[width=0.8\linewidth]{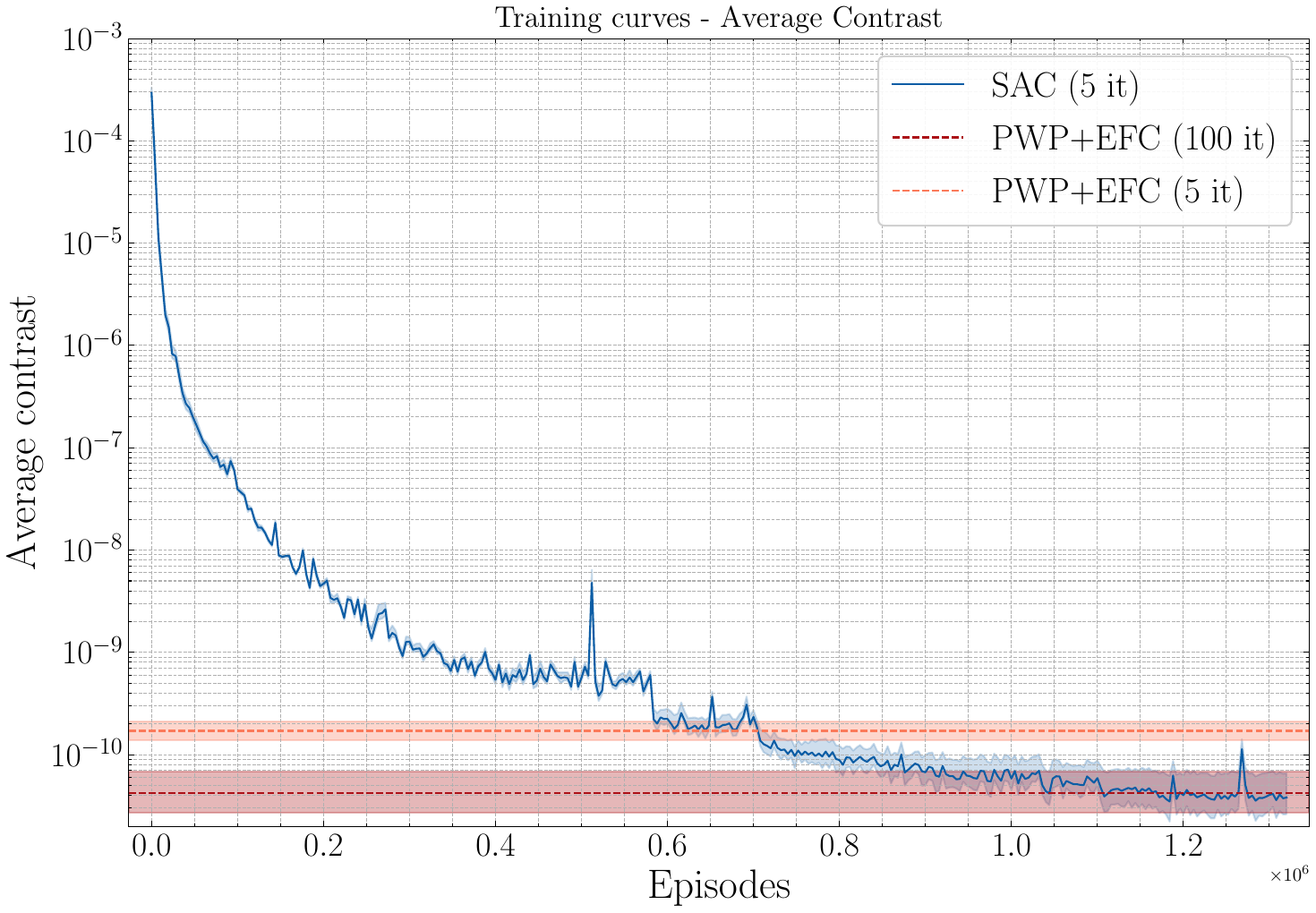}
    \caption{Evolution of the mean dark-hole contrast during SAC training.}
    \label{fig:sac}
\end{figure}

Figure~\ref{fig:sac} shows the evolution of the mean dark-hole contrast during training. The agent required approximately $1.2\times10^6$ episodes to reach its best performance. The first stages of training were the most efficient: within the first $3\times10^5$ interactions, the contrast decreased from about $10^{-4}$ to $10^{-9}$. After this initial improvement, further optimization became progressively harder. The last two orders of magnitude required most of the remaining training time, showing that each additional contrast gain becomes increasingly difficult at deeper contrast levels. With an appropriate learning-rate schedule, the agent was able to continue improving and eventually reached a contrast of approximately $4\times10^{-11}$.

\begin{figure}[h!]
    \centering
    \includegraphics[width=0.8\linewidth]{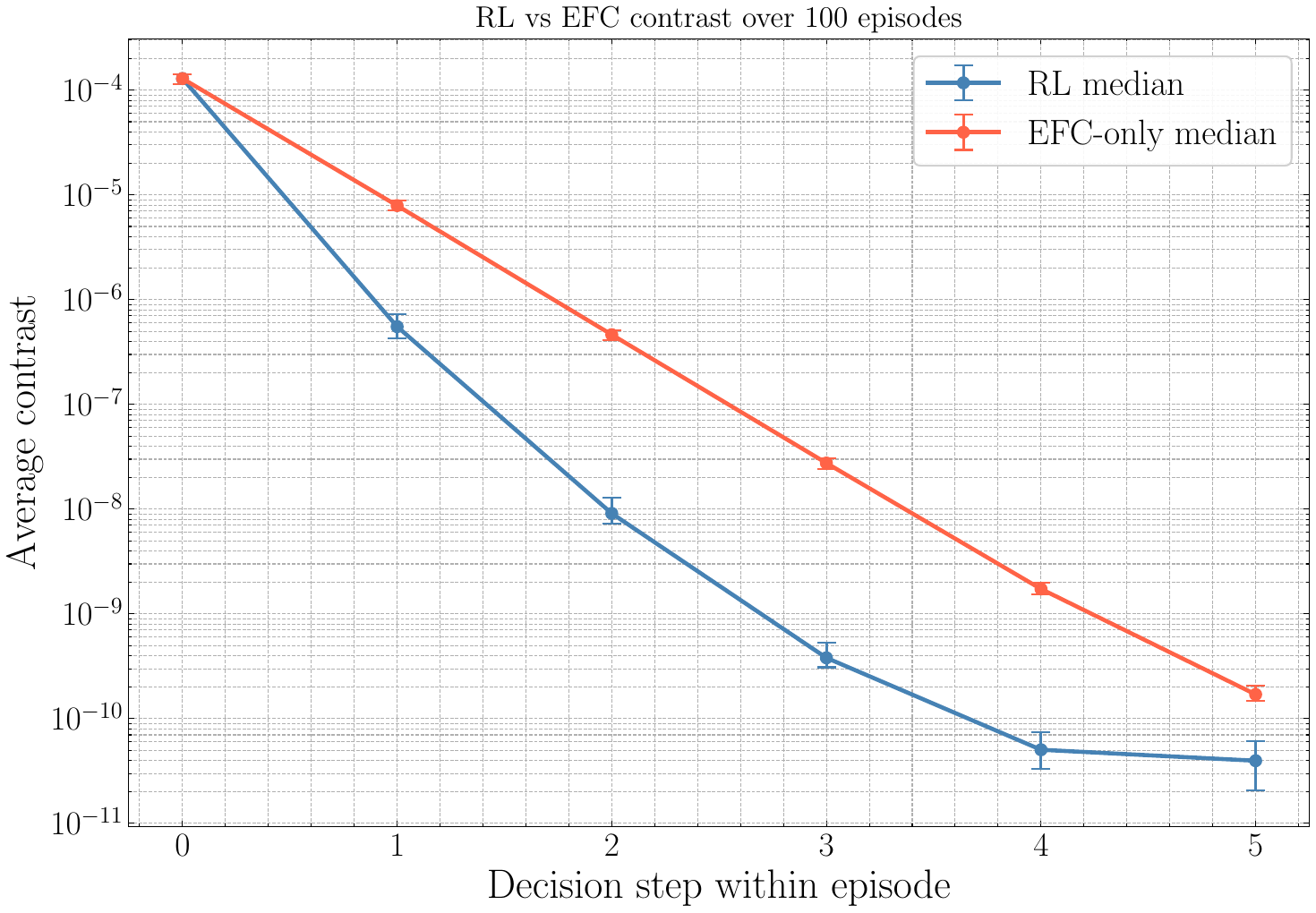}
    \caption{Evolution of the dark-hole contrast obtained with the RL controller and PWP+EFC over 100 independently sampled aberrations. Markers indicate the median, and vertical bars show the 25th--75th percentile interval.}
    \label{fig:efc_rl}
\end{figure}

Figure~\ref{fig:efc_rl} compares the contrast evolution obtained with the RL controller and the classical PWP+EFC approach over 100 independent episodes. For each correction step, the marker represents the median dark-hole contrast reached by each method, while the vertical bars indicate the interquartile range between the 25th and 75th percentiles.

The RL controller reaches deeper contrast levels in fewer iterations. By the fourth correction step, it achieves a contrast more than one order of magnitude lower than PWP+EFC. However, the final correction step appears more difficult for the learned policy: the improvement becomes marginal and, in some cases, the contrast slightly degrades. As a result, the final gap between RL and PWP+EFC is reduced, with PWP+EFC remaining within about one order of magnitude of the RL performance at the end of the episode.

The dispersion of the results also differs between the two methods. PWP+EFC shows a small interquartile range at all correction steps, indicating stable and consistent behavior across episodes. In contrast, the RL controller presents a larger spread, especially during the last two steps, suggesting that its performance becomes more sensitive near the deepest-contrast regime.

\subsection{Photon Noise}

\begin{figure}[h!]
\begin{center}
\begin{tabular}{c} 
\includegraphics[height=9.5cm]{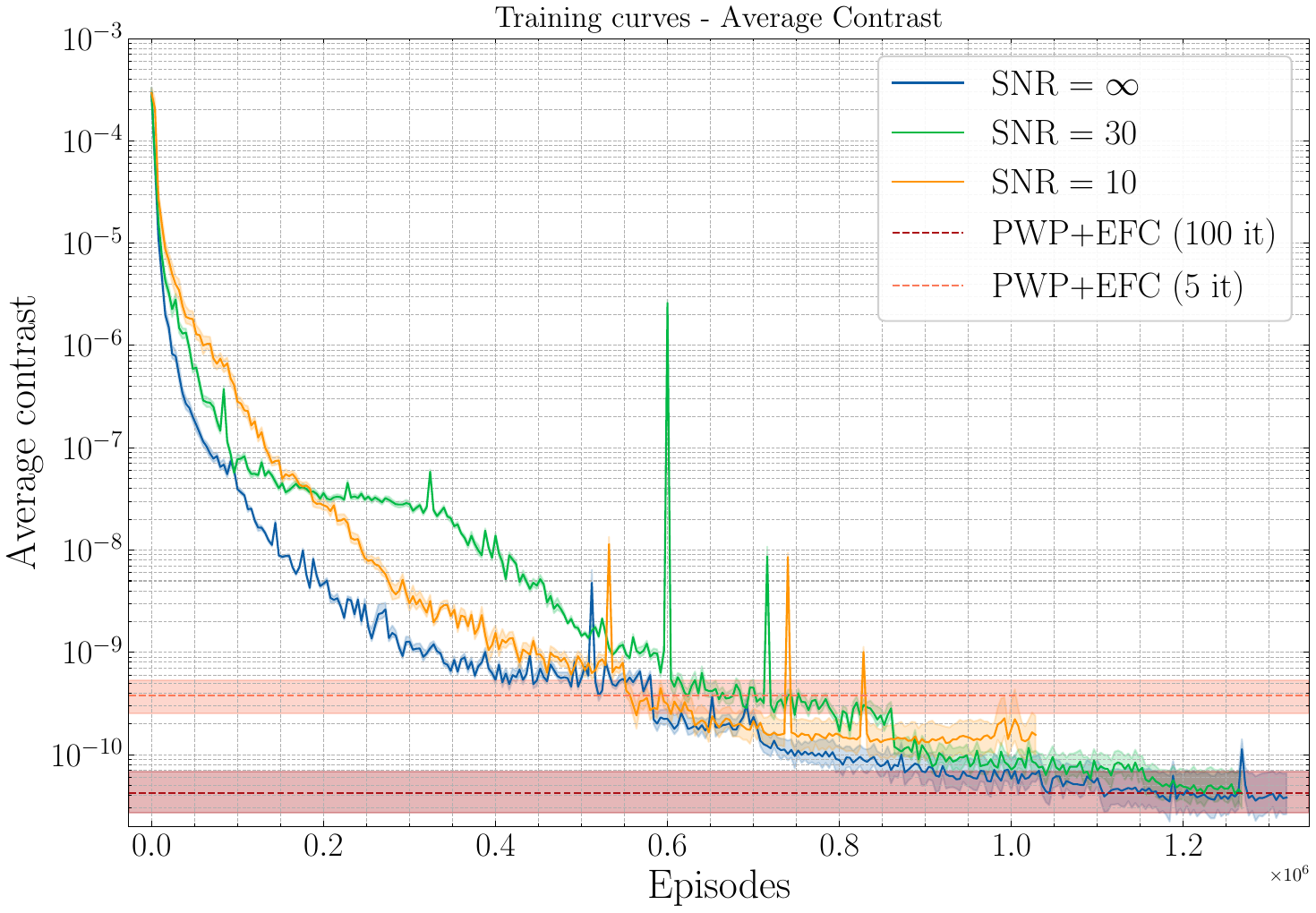}
\end{tabular}
\end{center}
\caption{\label{fig:snr} Training curve SAC}
\end{figure}

Figure~\ref{fig:snr} shows an initial evaluation of the learned policy under photon noise, comparing the noiseless case, denoted as ${\rm SNR}=\infty$, with noisy observations at ${\rm SNR}=30$ and ${\rm SNR}=10$. Overall, the performance degradation remains limited. For ${\rm SNR}=30$, the contrast evolution is close to the noiseless case, while for ${\rm SNR}=10$ the final contrast remains within less than one order of magnitude of the noiseless performance.

The main effect of photon noise is observed in the convergence behavior. Without noise, the agent reaches the deepest contrast levels more rapidly and follows a smoother correction path. In the noisy cases, the controller may require more steps to identify an effective correction direction, leading to a slower decrease of the dark-hole contrast. Nevertheless, these preliminary results suggest that the learned policy retains a significant part of its correction capability under moderate photon noise.

\section{CONCLUSION}

This work presented a proof of concept for reinforcement-learning-based dark-hole generation in a simplified high-contrast imaging simulation. The objective was to investigate whether a model-free controller could learn to control a deformable mirror directly from image-based feedback and generate a dark hole without relying on an explicit linearized optical model.

Achieving stable learning required several design choices, including reward optimization, hyperparameter tuning, observation-space design, and learning-rate management. Under the simplified simulation conditions considered here, the trained RL controller was able to create a dark hole within five correction steps and reach deep contrast levels. Compared with the classical PWP+EFC baseline, the RL controller reached lower contrast faster, achieving an improvement of more than one order of magnitude by the fourth correction step. An initial photon-noise study also showed limited degradation at moderate noise levels, suggesting that the learned policy retains part of its correction capability under noisy observations.

These results remain preliminary and were obtained under idealized simulation conditions. The next step is to progressively introduce more realistic effects and evaluate the robustness of the learned controller. Since PWP+EFC relies on a calibrated linear model, its performance is expected to become increasingly sensitive to model mismatch, temporal drifts, vibrations, misalignments, and other experimental uncertainties. In contrast, the model-free nature of RL may provide improved robustness if such effects are represented during training, for example through observation or domain randomization.

The long-term objective is experimental validation on the THD2 testbed at Observatoire de Paris--PSL \cite{Laginja2026THD2}. Three main challenges must be addressed to achieve this objective. First, the current framework must be extended from phase-only aberrations to mixed phase-and-amplitude aberrations. In this configuration, correction with a single deformable mirror requires restricting the dark hole to one side of the focal plane. Second, the number of controlled modes must be increased to represent realistic high-order wavefront errors, despite the resulting growth in the dimensionality of the action space and the associated learning difficulty. Finally, transferring the controller to the experimental system will require strategies such as transfer learning or fine-tuning to adapt the simulated policy to the characteristics and uncertainties of the testbed.

Addressing these challenges will be essential to determine whether reinforcement learning can become a robust alternative or complement to model-based wavefront-control strategies for future exoplanet imaging missions.

\acknowledgments

The Ph.D. work of Manuela Castañeda-Medina is co-funded by CNES and ONERA. This work was carried out as part of a doctoral research project jointly hosted by ONERA and the Observatoire de Paris--PSL. This project has received funding from the European Research Council (ERC) under the European Union’s Horizon Europe research and innovation programme under the grant agreement \#101230218.

\noindent \textit{Software} - This study made use of the following Python packages: \texttt{Asterix}\cite{MazoyerAsterixSimulator}, \texttt{Astropy}\cite{astropy_collaboration_2022}, \texttt{Matplotlib} \cite{hunter2007matplotlib}, \texttt{NumPy}\cite{harris2020NumPy}, \texttt{PyTorch}~\cite{paszke2019pytorch}, \texttt{SciPy}~\cite{virtanen2020scipy}, and \texttt{Stable-Baselines3}~\cite{raffin2021stablebaselines3}. OpenAI Codex was used to assist with code development, debugging, and refactoring, as well as with language editing and manuscript revision. All AI-assisted content and code was reviewed, tested, and validated by the authors. 

\bibliography{report} 
\bibliographystyle{spiebib} 

\end{document}